\documentclass[twocolumn,showpacs,preprintnumbers,amsmath,amssymb,prb,superscriptaddress]{revtex4}

\usepackage{graphicx}

\newcommand{\expct}[2]{\left\langle #1 \right\rangle_{#2}}

\newcommand{\C}{\mathcal{C}}
\newcommand{\TC}{T_{\C}\, }
\newcommand{\T}{\mathcal{T}}
\newcommand{\Exp}[1]{\text{e}^{#1}}

\begin{document}

\title{Transport properties of a molecular quantum dot coupled to
one-dimensional correlated electrons}
\author{S. Maier$^1$ and A. Komnik}
\affiliation{Institut f\"ur Theoretische Physik,
Ruprecht-Karls-Universit\"at Heidelberg,\\
 Philosophenweg 19, D-69120 Heidelberg, Germany}
\date{\today}

\begin{abstract}
 We analyze the transport properties of a quantum dot with a harmonic degree of
 freedom (Holstein phonon) coupled to interacting one-dimensional metallic
 leads. Using Tomonaga-Luttinger model
 to describe the interacting leads we construct the generating function of the full counting
 statistics (FCS) for a specific constellation of system
 parameters and give explicit expression for the cumulant generating function. In the resonant case we find the lowest order correction to the
 current to be negative and divergent when source-drain voltage approaches the phonon
 frequency. Via a diagram resummation procedure we show, that this divergencies can be repealed. On the contrary, in the off-resonant case the lowest order correction remains
 finite. This effect can be traced back to the strongly non-monotonic
 behaviour of the bare transmission coefficient (without phonon)
 with respect to the dot level energy. We calculate corrections to
 the noise power as well and discuss possible experimental implications of this
 phenomenon.
\end{abstract}


\maketitle

Quantum impurities are often used to model ultrasmall quantum dots
and contacted molecules. One of the most fundamental setups is the
Anderson impurity model taking into account the local Coulomb
interaction \cite{Anderson1961}. Especially in the case of
contacted molecules, however, an explicit consideration of
coupling to vibrational degrees of freedom is also desirable. This
is accomplished by the Anderson-Holstein model
\cite{Holstein1959,Hewson2002}. Even under equilibrium conditions
and in the absence of Coulomb repulsion it turns out to possess
interesting properties. Recently a number of successful electron
transport experiments carried out on contacted molecules revealed
some very interesting details
\cite{Zhitenev2002,Qiu2004,Yu2004,Pasupathy2005,Sapmaz2006,djukic05,Smit2002}.
Probably the most prominent of them is the different behavior of
the molecule conductance which can grow or decline as soon as the
applied voltage gets larger than the phonon
frequency.\cite{Vega2006,Paulsson2005,Mii2003,Egger2008,Schmidt2009,Avriller2009,Haupt2009}
This phenomenon can be understood as follows. At zero temperature
and voltages the vibrational degrees of freedom can be safely
assumed to be frozen out and one effectively deals with the
(noninteracting) resonant level with some energy $\Delta$. The
spectral function of the quantum dot is a single Lorentzian with
some width $\Gamma$ (which is related to the hybridization of the
dot level with the electrode) centered around $\Delta$. For the
large initial transmittance of the system $\Delta$ should lie in
between the chemical potentials of the contacting electrodes. On
the opposite, for small transmittance $\Delta$ is below/above the
chemical potentials. The system is virtually insulating at
$|\Delta| \gg \Gamma$ because then the spectral weight around the
chemical potentials position, which is necessary for transmission,
is very small. When the phonon gets excited its spectral function
is known to develop equidistant sidebands
\cite{Mahan1991,Braig2003}. The central peak at $\Delta$ persists
but due to spectral weight redistribution its height diminishes.
Therefore the initially large transmission drops as soon as the
vibrational degrees of freedom can be excited. On the contrary,
due to the finite spectral weight in the sidebands the conductance
grows for the out-of-resonance $\Delta$. It turns out that in
general the crossover from enhanced to suppressed transmission
does not correspond to any universal parameter constellation apart
of the limiting cases of large/small $\Delta$
\cite{Egger2008,Schmidt2009}. Nonetheless it has been observed in
several
experiments\cite{Zhitenev2002,Qiu2004,Yu2004,Pasupathy2005,Sapmaz2006,djukic05,Smit2002}.

Thus far only noninteracting electrodes were considered. Given the
small dimensions of the corresponding devices it is very likely
that the electrodes might in fact possess genuine one-dimensional
geometry as far as the electronic degrees of freedom are
concerned. Alternatively one might conceive a device contacted by
e.~g. armchair carbon nanotubes, which are known to host
one-dimensional electrons. In these situations instead of
conventional Fermi liquids one deals with the Tomonaga-Luttinger
liquids (TLL). Their most prominent feature is the power-law
singularity of the local density of states in vicinity of the Fermi
edge. Among other things it results in complete suppression of
transmission in presence of impurities in the low energy sectors
leading to the zero bias anomaly (ZBA)
\cite{Kane1992,Furusaki1993}. As a result the transmission through
a featureless quantum dot coupled to two TLLs vanishes towards
small voltages and low temperatures
\cite{Furusaki1998,Kane1992,Nazarov2003a,Komnik2003a,Polyakov2003}.
The only exception is the perfect resonant setup when $\Delta=0$
and hybridizations with both electrodes are equal to each other.
Thus, contrary to the noninteracting electrodes, when the dot
transmission can smoothly interpolate between perfect and zero
transmission, in the TLL setup only two low-energy transmission
regimes are possible: either zero or unity. Applying the above
line of reasoning one might conclude that in the former case the
current through the system starts to flow only after voltage gets
larger than the phonon frequency. In the opposite case one would
expect that the conductance of an initially perfectly transmitting
dot would rapidly decrease beyond the threshold set by the phonon
frequency. The goal of our paper is to understand the details of
transport properties of such a setup and to quantify this
heuristic picture.

In order to proceed one needs a model which can equally good
describe the off-resonant as well as perfectly transmitting case.
While in the noninteracting case the resonant level model without
phonon is trivially solvable this is not the case any more for the
TLL with a generic interaction strength. Nonetheless, at one
particular parameter constellation the problem is exactly
solvable. We take advantage of this and analyze the transmission
properties of the system with the phonon using the perturbation
theory in the electron-phonon coupling.
%
%
%
%
%
%
We model the system by the following Hamiltonian,
\begin{eqnarray}
 H = H_{\rm{leads}} + H_{\rm{dot}} + H_{\rm{tunn}} + H_{\rm{int}} \, .
\end{eqnarray}
$H_{\rm{leads}}$ describes two (R/L, right/left) metallic leads in
the Luttinger liquid state kept at different chemical potentials
$\mu_L - \mu_R = V$ with $V$ being the bias voltage applied across
the dot (we use units in which $e=\hbar=k_B=1$ throughout), see
for example Eqs.~(2-4) of [\onlinecite{Komnik2003}]. Since we are
not interested in any spin or finite field related effects our dot
Hamiltonian is given by
\begin{eqnarray}
H_{\rm{dot}} = \Delta d^\dag d + \Omega c^\dag c + g d^\dag d
(c^\dag + c) \, ,
\end{eqnarray}
which describes a single fermionic level with energy $\Delta$
coupled with the amplitude $g$ to the local oscillator with
frequency $\Omega$. $d$, $d^\dag$ and $c,c^\dag$ are the
respective annihilation/creation operators for the electron and
oscillator. $H_{\rm{tunn}}$ is responsible for the particle
exchange between the electrodes and the dot,
\begin{eqnarray}
 H_{\rm{tunn}} = \sum_{\alpha = R,L} \gamma_\alpha \, \psi_\alpha^\dag(x=0) d +
 \mbox{H. c.} \,
\end{eqnarray}
where $\gamma_{R,L}$ are the (constant) tunneling amplitudes. The
tunneling is assumed to be taking place locally at $x=0$ in the
coordinates of the respective electrode. The last term is the
capacitive coupling of the electrodes to the dot
\begin{eqnarray}
H_{\rm{int}} = U_C \, d^\dag d \left[ \psi_L^\dag(0) \psi_L(0) +
\psi^\dag_R(0) \psi_R(0) \right] \, .
\end{eqnarray}
The present problem allows for an exact solution for $g =0$ and
Tomonaga-Luttinger interaction parameter $K=1/2$ while $U_C = 2
\pi v_F$. The emergent structure is that of the Majorana resonant
level model (MRLM) extended by the vibrational degree of freedom
\cite{Komnik2009}, in the case of the symmetric coupling $\gamma_L
= \gamma_R = \gamma$
\begin{eqnarray}                        \label{MRLM1}
 H = H_0[\xi,\eta] - i \, [\Delta + g (c^\dag + c)] \, a b +
 i \gamma \, b \, \xi(0) +\Omega c^\dag c \, .
\end{eqnarray}
The first part of the transformed Hamiltonian describes the
electrodes in terms of new Majorana fermions $\xi$ and $\eta$,
\begin{eqnarray}                      \label{H0prime}
 H_0 = i \int\, dx \, \Big[
 \eta(x) \partial_x \eta(x) + \xi(x) \partial_x \xi(x)
 + V \xi(x) \eta(x) \Big] \, .
\end{eqnarray}
$a = (d^\dag + d)/\sqrt{2}$ and $b=-i(d^\dag - d)/\sqrt{2}$ are
the respective Majorana operators for the dot fermion. In the
absence of the phonon the transport properties of the above system
are known even on the level of the full counting statistics (FCS)
\cite{AM,Kindermann2005,Komnik2005}. It is accomplished by
introduction of the counting field $\lambda$ into the Hamiltonian
making it explicitly time-dependent due opposite sings of the
counting field on the different Keldysh branches. In the model
this is expressed in the replacement of the tunnel Hamiltonian
$H_T =-i\sqrt{2}\gamma b\xi$ by the operator $T_\lambda =
-i\sqrt{2}\gamma b\left[\xi\cos(\lambda/2) -\eta\sin(\lambda/2)
\right]$. In absence of phonon-electron interaction the electric
current is given by\cite{Komnik2003}
\begin{eqnarray}
 I = G_0 \int d \omega \, (n_L - n_R) D(\omega) \, ,
\end{eqnarray}
where $G_0 = 2 e^2/h$ is the conductance quantum, $n_{L,R}$ are
the Fermi distribution functions in the respective electrode and
\begin{eqnarray}                \label{trans}
 D(\omega) = \frac{\omega^2 \Gamma^2}{(\omega^2 - \Delta^2)^2 +
 \omega^2 \Gamma^2}
\end{eqnarray}
is an effective transmission coefficient for the new fermions. Its
nonmonotonic behavior around $\omega =0$ for zero and finite
$\Delta$ is the reason for the full transmission suppression in
the off-resonant case
\cite{Komnik2003a,Nazarov2003a,Polyakov2003}. The system under
consideration is not the only one with the effective transmission
coefficient of the form (\ref{trans}). It is also encountered in
the exact analytic solution of the two-terminal Kondo model at the
Toulouse point\cite{Schiller1998} as well as in the transmission
coefficient for the physical electrons in the parallel double
quantum dot setup\cite{Kubala2003,Dahlhaus2010}. The
generalization of the phenomena discussed below to these setups is
thus straightforward.

The cumulant generating function (CGF) for the system at $g\neq 0$
factorizes as $\chi(\lambda) = \chi_0(\lambda) \chi_g(\lambda)$,
where $\chi_0(\lambda)$ is the CGF calculated in
[\onlinecite{AM,Komnik2005,Kindermann2005}]. The correction due to
electron-phonon coupling is found from
\begin{gather}
 \ln \chi_g\left(\lambda\right)=\expct{\Exp{-g\int_\C d\tau
 \left[c^\dagger\left(\tau\right)+c\left(\tau\right)\right]a
 \left(\tau\right)b\left(\tau\right)}}{\lambda}
\end{gather}
where the expectation value is taken with respect to the operator
$H_0[\xi,\eta] - i \, \Delta \, a b +\Omega c^\dag c + T_\lambda$.
The time integration runs along the full Keldysh contour ${\cal
C}$ and the superscripts $k,l=\pm$ refer to the backward/forward
propagating branches. In the leading order perturbation theory
there are two different contributions: (i) the one generated by
the `tadpole' diagram
\begin{multline}
 \ln\chi_{g,1}^{(1)} = -i \frac{g^2}{2}\sum_{k,l=\pm}(kl)
 \int_\C d\tau_1 d\tau_2 G_{ph}^{kl}\left(\tau_{12}\right)\\
 \times \lim_{\epsilon \searrow 0}D_{ab}^{kk}\left(-k\epsilon\right)D_{ab}^{ll}\left(-l\epsilon\right) \,
 ;
\end{multline}
(ii) the one given by the `shell' diagram
\begin{multline}
  \ln\chi_{g,2}^{(1)} = -i \frac{g^2}{2}\sum_{k,l=\pm}kl\int_\C d\tau_1
  d\tau_2 G_{ph}^{kl}\left(\tau_{12}\right) \\ \times \biggl[
  D_{ab}^{kl}\left(\tau_{12}\right)D_{ab}^{lk}\left(-\tau_{12}\right)-
  D_{aa}^{kl}\left(\tau_{12}\right)D_{bb}^{lk}\left(-\tau_{12}\right)\biggr]
  \, .
\end{multline}
Here, we introduced the phonon Keldysh Green's function (GF)
$G_{ph}\left(\tau_{12}\right) = -i\expct{\TC c\left(\tau_1\right)
c^\dag\left(\tau_2\right)}{}$, the Keldysh functions of the
quantum dot $D_{\alpha\beta}\left(\tau_{12}\right) = -i\expct{\TC
\alpha\left(\tau_1\right)\beta\left(\tau_2\right)}{\lambda}$ where
$\alpha$ and $\beta$ take values in the set $\{a,b\}$ and
$\tau_{12}=\tau_1-\tau_2$. At this point it is convenient to
transform to energy variables
\begin{gather}
  \ln\chi_{g,2}^{(1)} =  \frac{g^2}{2}\sum_{k,l=\pm}\left(kl\right)
   \int \frac{d\omega}{2\pi} G_{ph}^{kl}\left(\omega\right) \pi^{kl}\left(\omega\right)
\end{gather}
and introduce generalized (Keldysh) polarization loops
\begin{multline}
\pi^{kl}\left(\omega\right)=-i \int\frac{dy}{2\pi}\biggl[D_{ab}^{kl}
\left(y+\omega\right)D_{ab}^{lk}\left(y\right)\\-D_{aa}^{kl}\left(y+\omega\right)
D_{bb}^{lk}\left(y\right)\biggr]
\end{multline}
In the resonant case ($\Delta=0$) the mixed dot Keldysh GFs
($\alpha\neq \beta$) vanish identically  and
$D_{aa}\left(\omega\right)$ becomes diagonal. Hence, we are left
with the calculation of the (anti-)time-ordered components of the
polarization loop. In the zero temperature limit the first order
correction to the full counting statistics can be calculated
exactly. One finds
\begin{gather*}
 \ln\chi_g^{\left(1\right)}=-\frac{\T g^2}{2\pi}\frac{\Exp{i\frac{\lambda}{2}}
 \Gamma\tan^{-1}\left(\frac{V}{\Gamma}\Exp{-i\frac{\lambda}{2}}\right)-\Omega
 \tanh^{-1}\left(\frac{V}{\Omega}\right)}{\Gamma^2 +\Omega^2+\Gamma^2
 \left(\Exp{i\lambda}-1\right)}
\end{gather*}
Thus the correction to the transport current has the following
form,
\begin{multline}
 I^{\left(1\right)} = \frac{g^2}{2\pi}\Gamma^2\biggl[\frac{V}{
 \left(V^2+\Gamma^2\right)\left(\Gamma^2+\Omega^2\right)} \\
 +\frac{\left(\Gamma^2-\Omega^2\right)\tan^{-1}\left(V/\Gamma\right)
 - 2\Gamma\Omega\,\tanh^{-1}\left(V/\Omega\right)}{\Gamma\left(\Gamma^2
 +\Omega^2\right)^2}\biggr]
\end{multline}
Expressions for noise and higher cumulants can easily be
determined by taking higher order derivatives of the CGF with
respect to $\lambda$ and setting it to zero afterwards. The
calculation of the correction to the full statistics at finite
temperature is quite involved. It is, in fact, more convenient to
set up explicit expressions for the individual cumulants before
the energy integration.
\begin{figure}[h]
\includegraphics[width=8.5cm]{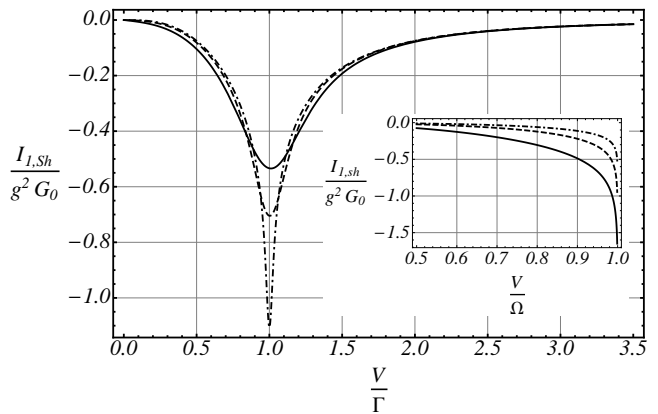}
\caption{\emph{Main graph:} first order current correction at
finite temperature. The parameters are $\Omega/\Gamma=1$ and
$T/\Gamma= 0.1, 0.05, 0.01$ (solid, dashed , dashed-dotted). \emph{Inset:}
first order current corrections in case of zero temperature for
$\Gamma/\Omega=1,2,3$ (solid, dashed, dashed-dotted)}
\label{fig1}
\end{figure}
A remarkable property is the absence of any sharp energy threshold
usually found in phonon affected transport between uncorrelated
electrodes. Those are usually attributed to the onset of inelastic
processes \cite{Vega2006,Egger2008,Schmidt2009}, when individual
electrons may loose/gain energy $\Omega$ during tunneling between
the leads at $V \ge \Omega$.
In the present case of MRLM the situation is different. The
Majorana fermions describe the collective excitations of the TLL
state in the leads (kinks and antikinks), rather then individual
electrons. That is why even at $T=0$ there is no sharp energy
threshold. Another feature is the logarithmic divergency for
$V\rightarrow \Omega$, which does not survive at finite
temperatures, see Fig.~\ref{fig1}. It can be attributed to the
non-monotonic behavior of the effective transmission coefficient
(\ref{trans}) at zero and finite off-set $\Delta$. The
transmission for the quasiparticles drops immediately to zero as
soon as $\Delta$ becomes off-resonant. This is exactly what occurs
when the harmonic degree of freedom is excited -- the coupling $g$
generates an effective non-zero offset. In this voltage regime the
problem becomes non-perturbative in the electron-phonon coupling
and therefore requires analysis by more advanced methods. One
possible route would be an RPA-like diagram resummation.  To
perform this program, we express the CGF in terms of the adiabatic
potential \cite{AndersonFCS}, which itself can be rewritten as a
function of the exact $D_{bb}$ Keldysh GF only. It is approximated
by the RPA-type summation (see Fig.~\ref{fig:RPA} for the
diagrammatic structure).
\begin{figure}[h]
\centering
\includegraphics[width=8.5cm]{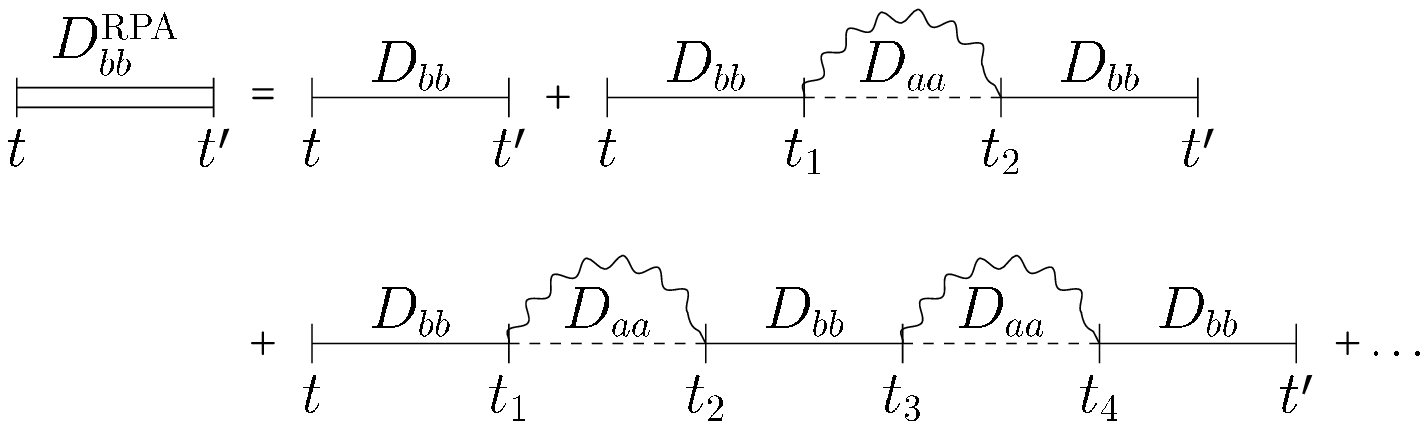}
\caption{Summation of an infinite series of diagramms. $
D_{bb}$,
$D_{aa}$  represents the Keldyshfunction of the $b$ resp. $a$
Majorana fermions as defined before with $\lambda$ set so zero and the wiggly lines represent the phonon
propagators} \label{fig:RPA}
\end{figure}
Interestingly, within this approximation one finds for the CGF the
Levitov-Lesovik formula\cite{lll}
\begin{gather}
\ln\chi^{\text{RPA}}\left(\lambda\right) = {\cal T}
\int\frac{d\omega}{2\pi}\ln\left[1+D_0^{\text{eff}}\left(\omega\right)\left(\Exp{
i\lambda}-1\right)\left(n_L -n_R\right)\right]
\end{gather}
with the effective transmission coefficient
\begin{gather}
 D_0^{\text{eff}}\left(\omega\right)=
 \frac{\Gamma^2\left(\omega^2-\Omega^2\right)^2}{g^4 \omega^2 -2 g^2
 \omega^2\left(\omega^2-\Omega^2\right)+\left(\omega^2+\Gamma^2\right)
 \left(\omega^2-\Omega^2\right)^2}
\end{gather}
which has a three maxima (perfect transmission) at $\omega = 0,
\pm \sqrt{g^2+\Omega^2}$ and minima (perfect reflection) at
$\omega = \pm \Omega$.
\begin{figure}[h]
\includegraphics[width=8.5cm]{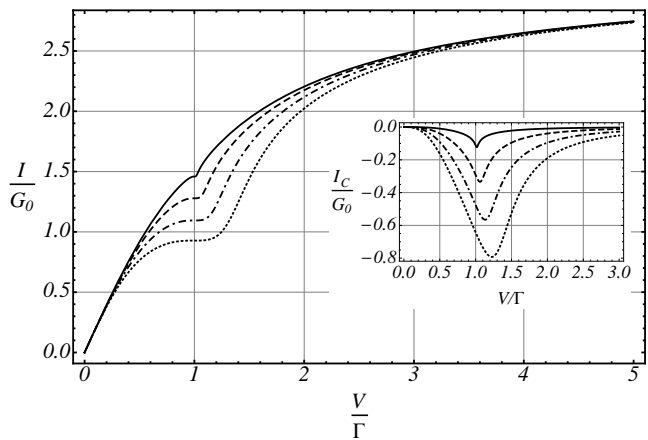}
\caption{\emph{Main graph:} Full current in RPA approximation. The
parameters are $\Omega/\Gamma=1$ and $g/\Gamma= 0.25, 0.5, 0.75 ,
1$ (solid, dashed , dashed-dotted, dotted lines, respectively).
\emph{Inset:} the current correction in RPA approximation
$\Omega/\Gamma=1$ and $g/\Gamma= 0.25, 0.5, 0.75, 1$ (solid,
dashed, dashed-dotted and dotted lines)} \label{fig:currentrpa}
\end{figure}

In Figs.~\ref{fig:currentrpa} and \ref{fig:noiserpa} the current
and noise in our approximation are depicted. One observes that for
small voltages the current increases nearly linear. For voltages
near $\Omega$ one finds a plateau, e.~g. the current enhancement
is suppressed by the electron-phonon interaction. This feature
does not occur exactly at $V=\Omega$ (see
Fig.~\ref{fig:currentrpa}). One finds that the maximal reduction
of the current is at $V=\Omega\sqrt{1+\frac{g^2}{2\Omega^2}}$ or
if one assumes $\Omega \gg g$, $V \approx
\Omega+\frac{1}{4}\frac{g^2}{\Omega}$. This kind of a shift by
$\frac{g^2}{\Omega}$ is one normally produced by a polaron
(Lang-Firsov) transformation\cite{Mahan1991}.

\begin{figure}[h]
\includegraphics[width=8.5cm]{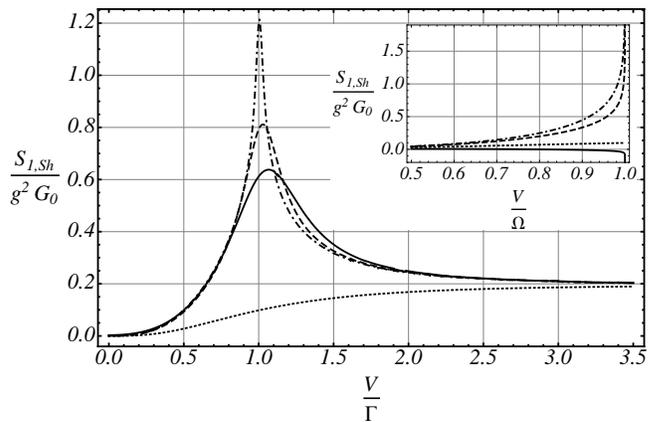}
\caption{\emph{Main graph:} $g^2$ correction to the shot noise at
finite temperatures. The plotting parameters are $\Omega/\Gamma=1$
and $T/\Gamma= 0.1, 0.05, 0.01, 0$ (solid, dashed, dashed-dotted,
dotted lines respectively). \emph{Inset:} $g^2$ noise corrections
in case of zero temperature. The plotting parameters are
$\Gamma/\Omega=1,2,3$ (solid, dashed, dashed-dotted lines)}
\label{fig2}
\end{figure}
\begin{figure}[h]
\includegraphics[width=8.5cm]{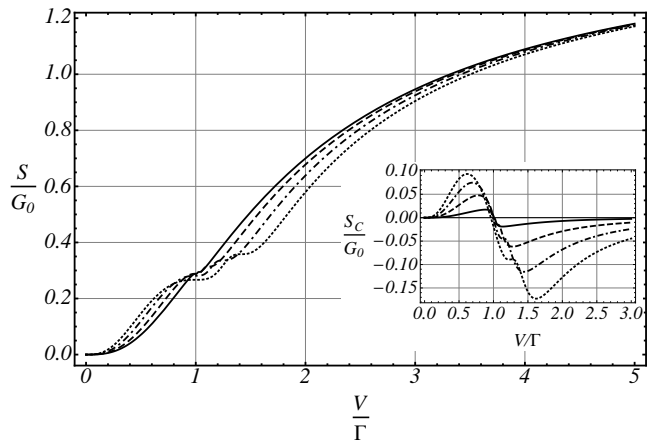}
\caption{\emph{Main graph:} Shot noise in RPA approximation. The
parameters are $\Omega/\Gamma=1$ and $g/\Gamma= 0.25, 0.5, 0.75,
1$ (solid, dashed , dashed-dotted and dotted lines). \emph{Inset:}
noise correction in RPA approximation for $\Omega/\Gamma=1$ and
$g/\Gamma= 0.25, 0.5, 0.75, 1$ (solid, dashed, dashed-dotted and
dotted lines).} \label{fig:noiserpa}
\end{figure}

The behavior of the shot noise turns out to be even more
interesting, see Fig.~\ref{fig2}. In case of zero temperature, to
the lowest order in $g$ one again finds a log-divergent
contribution for voltages approaching the phonon frequency.
However, for a special configuration $\Gamma=\Omega$ the
singularity cancels out and the shot noise interpolates in a
regular way between small and large voltage limits. For $\Omega >
\Gamma$ the correction to the shot noise develops a new feature:
it changes sign, see Fig.~\ref{fig2} inset. Just as in the case
of the transport current, the resummation heals the singularity,
see Fig.~\ref{fig:noiserpa}.

There is an important difference in the behavior of the noise and
current corrections in the high voltage limit. While the latter
tends to vanish the former remains finite. It can be understood as
follows. As already discussed in the introduction, for the current
only the overall spectral density between the chemical potentials
in the leads is essential. For $V \gg \Omega$ the phonon
excitations are less effective in squeezing the spectral weight
beyond the voltage window, so the current should approach the
unitary value already in the zeroth order in $g$. For the noise,
however, the rate of phonon (de)excitation is important. In the
limit $V \rightarrow \infty$ it achieves its maximal value,
thereby generating a finite contribution seen in Fig.~\ref{fig2}.
However, this feature isn't observed after the diagram subsummation procedure.

We now turn to the off-resonant system $\Delta\neq 0$. It is again
convenient to perform the $\lambda$ differentiations first and
integrate over energy afterwards. Here two different regimes of
weak $|\Delta| < \Gamma/2$ and strong detuning $|\Delta| >
\Gamma/2$ emerge.\cite{Komnik2003a,Komnik2009}

In the following we consider the case of weak detuning. Then, the
tadpole contribution yields
\begin{gather}
 I^{\left(1\right)}_{\text{tadpole}} =\frac{\Gamma\Delta}{2\pi\Omega}
 \frac{\ln\left(\frac{V^2+\Omega_+^2}{V^2+\Omega_-^2}\right)}
 {\Omega_-^2-\Omega_+^2}\partial_\Delta\sum_\pm\frac{\pm\Gamma^2
 \Omega_\pm \tan^{-1}\left(\frac{V}{\Omega_\pm}\right)}{\Omega_+^2-\Omega_-^2}
\end{gather}
where we have defined $\Omega_\pm = \Gamma/2 \pm
\sqrt{\left(\Gamma^2 / 4-\Delta^2\right)}$.
\begin{figure}[h]
\includegraphics[width=8.5cm]{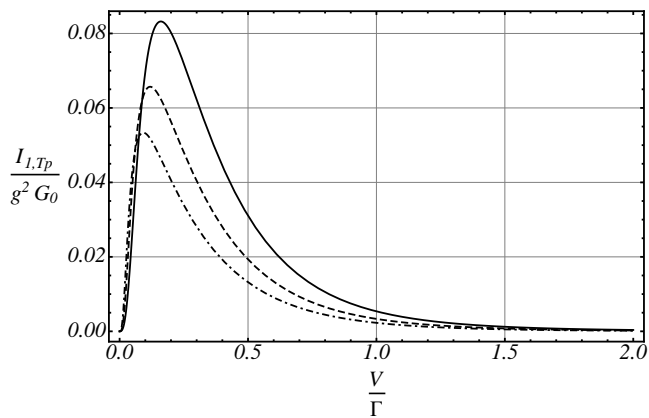}
\caption{{Tadpole correction to the current. The parameters of the
plots are $\Omega/\Gamma=10$ and $\Delta/\Gamma$ takes the values
$0.25, 0.2, 0.17$  (solid, dashed and dot-dashed lines,
respectively)}} \label{fig3}
\end{figure}
The corresponding current correction is plotted in
Fig.~\ref{fig3}.  Just as in the case of transport between two
uncorrelated electrodes\cite{Schmidt2009}, its simple functional
dependence on the phonon frequency hints at its dominance in the
(Born-Oppenheimer) limit
 of the very `slow' phonon\cite{Gogolin2002,Riwar2009}. The parameter
$\Gamma$ mainly determines the width of the peak and $\Delta$ its
position. Because of the simple form of this contribution, we will
omit it in further illustrations.

\begin{figure}[h]
\includegraphics[width=8.5cm]{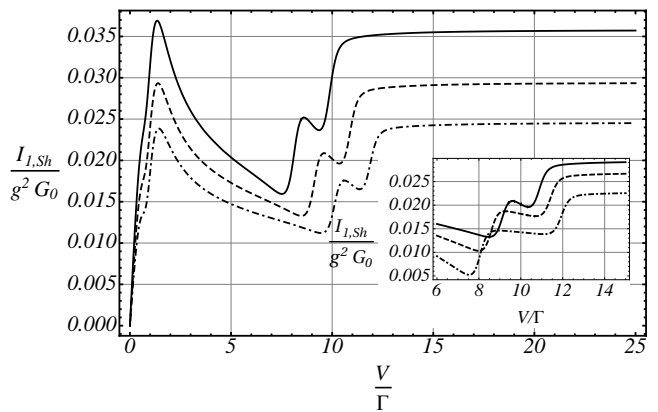}
\caption{{\emph{Main graph:} shell diagram correction to the
current. The parameters of the plots are  $\Delta/\Gamma = 1$ and
$\Omega/\Gamma$ takes the values $9,10,11$  (solid, dashed,
dashed-dotted)}. \emph{Inset:} double step like feature for the
parameters $\Omega/\Gamma=10$ and $\Delta/\Gamma=1,1.5,2$ (solid,
dashed, dashed-dotted). In both, the main graph and the inset, we chose strong detuning, because in this limit electron-phonon interaction effects are more pronounced.} \label{fig4}
\end{figure}

The contributions from the shell diagram are more sophisticated:
\begin{gather}
 I_{\text{shell}}^{(1)} = \frac{\left(g\Gamma\right)^2}{2}
 \int_{-V}^{V}\frac{dy}{2\pi} \left[j_1\left(y\right) + j_2\left(y\right)
 + j_3\left(y\right)\right] \, ,
\end{gather}
where we have
\begin{gather*}
 j_1\left(y\right) = -\sum_\pm \frac{\Delta^2\left(y^2-\Delta^2\right)^2}
 {D_0^2\left(y\right)} \frac{y^2+\left(y\pm \Omega\right)^2 +2y\left(y\pm \Omega\right)}
 {D_0\left(y\pm\Omega\right)}\\
 \times\left[1+\Theta\left(-V-y\mp\Omega\right)\right]\,
\end{gather*}
\begin{gather*}
 j_2\left(y\right) = -\sum_\pm\biggl[ \frac{y\Delta^2\left(\Delta^2-
 \left(y\pm\Omega\right)^2\right)+y\Gamma^2\left(y\pm\Omega\right)}
 {D_0\left(y\pm\Omega\right)} \\+ \frac{\left(y^2+\Delta^2\right)
 \left(y\pm\Omega\right)\left[\left(y+\Omega\right)^2-\Delta^2\right]}
 {D_0\left(y\pm\Omega\right)}\biggr]\frac{y\left(y^2-\Delta^2\right)}
 {D_0^2\left(y\right)} \, ,
\end{gather*}
where $D_0\left(y\right) =
\left(y^2-\Delta^2\right)^2+y^2\Gamma^2$. The expression von $j_3$
is quite lengthy, see Appendix for details.
Similar to the tadpole contribution this part is also always
positive.  In contrast to the resonant case, the corrections
remain finite for every voltage. For $V$ close to the phonon
frequency we observe a double step like feature instead. This can
be attributed to the double-maximum behavior of the transmission
coefficient (\ref{trans}), the denominator $D_0(y)$ of which
enters all relevant Green's functions and thereby affects the
voltage dependence. The linear scaling of the distance between the
dips is shown in Fig.~\ref{fig4}.
\newline
To conclude, we investigated the interacting resonant level model
in presence of a harmonic degree of freedom coupled to the quantum
dot. We observe that in the resonant case, where the system is
initially perfectly transmitting, finite electron-phonon coupling
leads to negative corrections to the current. In the zero
temperature limit we identified a strongly non-perturbative regime
where the current correction is log-divergent and performed an
RPA-like resummation of divergent diagram contributions, which
turned out to produce a plateau-like feature in the full
current-voltage characteristics of the system. We believe that
this behaviour is generic in all setups with TLL electrodes also
beyond the chosen parameter constellation. Single wall carbon
nanotubes (SWNTs) are known to be typical realizations of the TLL
electronic
state.\cite{egger_gogolin_prl,kane_fisher,bockrath,Yao1999}
Therefore we expect the above strong conductance suppression
phenomenon to be observable in experiments on molecular quantum
dots coupled to SWNTs. In the opposite off-resonant case, when the
system without the phonon has zero conductance, we observe
conductance enhancement due to electron-phonon interaction. For
voltages comparable to phonon frequency we find a double-step like
feature in the lowest order perturbation expansion in
electron-phonon coupling. Contrary to the resonant case no
singularities are observed.



The authors would like to thank T.~L.~Schmidt for many interesting
discussions. The financial support was provided by the DFG under
grant No.~KO~2235/3 and by the Kompetenznetz ``Funktionelle
Nanostrukturen III'' of the Baden-W\"urttemberg Stiftung
(Germany).

\appendix
\section{Tadpole contribution in case of strong detuning}
In case of strong detuning $\Delta^2 > \frac{\Gamma^2}{4}$ one finds for the static contribution the expression
\begin{gather}
I_{\text{static}}=\frac{4}{\pi\Omega}\sum_{\pm}\frac{\tanh^{-1}\left[\frac{\pm 2V-I\Gamma}{\sqrt{4\Delta^2-\Gamma^2}}\right]}{\sqrt{4\Delta^2-\Gamma^2}}\\ \times\partial_\Gamma \left[\frac{\sum\limits_{\pm} \pm \tanh^{-1}\left[\frac{\pm 2V -I\Gamma}{\sqrt{4\Delta^2-\Gamma^2}}\right]}{\sqrt{4\Delta^2-\Gamma^2}}\right]\text{.}
\end{gather}
\section{Explicit expression of the $j_3$ current contribution }
With the definitions
\begin{align*}
 &f_{rp}:=\left(\omega+r\Omega\right)^2-r\Omega_p^2 \\
&g_p:=\Omega^4 + 2\Omega^2\left(\Omega_p^2-\omega^2\right)+\left(\omega^2+\Omega_p^2\right)^2\\
&h_r:=\ln\frac{\left(\omega-r\Omega\right)^2-V^2}{\Gamma^2}\\
&k_p:=\ln\frac{V^2+\Omega_p^2}{\Gamma^2}
\end{align*}
where the indices $r,p$ run over $\left\{\pm\right\}$ together with
\begin{gather*}
 h_1=\sum_{r,p=\pm}\Biggl[ \frac{\pi p\Omega\left(\omega^2-\Omega^2+\Omega_+\Omega_-\right)}{\left\vert \Omega_p\right\vert\left[\Omega^2+\left(i\omega+\Omega_+\right)^2\right]\left[\Omega^2+\left(i\omega+\Omega_-\right)^2\right]}\\+\frac{p f_{rp}h_r +\omega\Omega k_p}{g_p} + \frac{\pi p\Omega\left\vert\Omega_p\right\vert\left(\Omega^2-\omega^2+\Omega_p^2\right)}{g_p}\Biggr]
\\h_2=\sum_{r,p=\pm}\Biggl[\frac{i\pi p\Omega\Omega_p\left(i\omega+\Omega_p\right)}{\left[\Omega^2+\left(i\omega+\Omega_+\right)^2\right]\left[\Omega^2+\left(i\omega+\Omega_-\right)^2\right]} \\+ \frac{1}{2}\frac{\pi\left\vert\Omega_p\right\vert}{\left(\omega-\Omega\right)^2+\Omega_p^2}+\frac{\left(\omega+r\Omega\right)\left(h_r+r k_p\right)}{\left[\left(\omega-r\Omega\right)^2+\Omega_p^2\right]}\Biggr]\frac{4}{\Omega_+^2-\Omega_-^2}
\\h_3=\sum_{p=\pm}\Biggl[\frac{-2\pi p\Omega\left(\Omega^2-\omega^2+\Omega_p^2\right)}{\left\vert\Omega_p\right\vert g_p\left(\Omega^2_+-\Omega^2_-\right)} +\frac{p f_{p+}f_{p-}h_p}{g_+g_-}\\-\frac{p\omega\Omega k_p}{g_p\left(\Omega_+^2-\Omega_-^2\right)}\Biggr]
\end{gather*}
one finds for $j_3$
\begin{multline}
j_3\left(\omega\right) = \frac{\left(\omega^2-\Delta^2\right) \omega\Delta^2\Gamma^3}{\left(4\pi\right)D_0\left(\omega\right)^2}\Biggl[h_1\left(\omega\right) +\omega h_2\left(\omega\right)+ \omega^2 h_3\left(\omega\right)\Biggr]
\end{multline}

\bibliography{LL_Phonon}
\end{document}